\documentclass{PoS}

\usepackage{epsfig} \newcommand{\BK}{\(B_K\)\ }

\PoS{PoS(LAT2005)346}

\title{Preliminary Study of \BK on 2+1 flavor DWF lattices from QCDOC}
\ShortTitle{Preliminary Study of \BK}
 
\author{\speaker{Saul~D.~Cohen}\thanks{for the RBC and UKQCD
	Collaborations} \\ Columbia University \\ E-mail:
	\email{sdcohen@phys.columbia.edu}}

\abstract{I present some preliminary calculations of \BK on 2+1 flavor
domain-wall fermion lattices from the QCDOC, including a set of
\(16^3 \times 32 \times 8\) lattices with \(a^{-1}\) near 1.6 GeV.
Although a final result awaits the production of a much longer run,
I will compare this preliminary value to previous results.}

\FullConference{XXIIIrd International Symposium on Lattice Field
	Theory \\ 25-30 July 2005 \\ Trinity College, Dublin, Ireland}

\begin{document}

\section{Introduction}

The study of hadrons allows experimental access to a wide variety of
phenomena. However, at low energies QCD does not provide easy
theoretical tools for parsing experimental results into measurements
of fundamental parameters. Here, experimentalists may turn to lattice
QCD to provide first-principles calculations of the factors required
to connect experiment with theory.

In this proceeding, I will describe preliminary calculations of \BK on
2+1 flavor domain-wall fermion lattices.

\section{The Kaon Bag Parameter}

In the Standard Model, there are two sources of CP violation: direct
and indirect.  In indirect CP violation, decay into a disallowed
number of pions is due a small amount of mixing between the \(K^0\)
and the \(\overline{K^0}\). In order to use kaon data to constrain
the unitarity triangle, we must disentangle the magnitude of kaon
mixing, parametrized by the kaon bag parameter \BK.

It is defined by
\begin{equation}
B_K = \frac{\langle\overline{K^0}|{\cal O}^{\Delta S=2}_{LL}|K^0\rangle}{\frac{8}{3}f_K^2M_K^2},
\end{equation}
where \(M_K\) is the mass of the neutral kaon, \(f_K\) is the decay
constant of the kaon (given by its coupling to the axial current), and
\(O^{\Delta S=2}_{LL}  = (\overline{s}d)_L(\overline{s}d)_L\) is a
four-quark operator coupling to left-handed quarks that changes
strangeness by 2.

\section{Challenges}

In the chiral limit, \BK contains only the operator of the form
\(VV+AA\):
\begin{equation}
{\cal{O}}^{\Delta S = 2}_{VV+AA} = (\overline{s} \gamma_\mu d)(\overline{s} \gamma_\mu d) + (\overline{s} \gamma_\mu \gamma_5 d)(\overline{s} \gamma_\mu \gamma_5 d),
\end{equation}
which may be derived from \(L = (V-A)\) and parity symmetry, which
forbids \(VA\) and \(AV\) terms. However, on the lattice we typically
do not have the luxury of chiral symmetry; in this case, we must
consider mixing with other chiral structures.

Even worse, it may be shown in chiral perturbation theory that these
operators are not proportional to mass in the chiral limit, while the
desired operator is. If significant mixing occurs, we will see almost
nothing of the continuum operator we desire.

Domain-wall fermions offer a solution by allowing us to control the
amount of chiral symmetry breaking by varying the size of the fifth
dimension. The results of symmetry breaking are easily treated by the
adding a residual mass. It is crucial that our residual mass be kept
small, since the unwanted off-chirality terms will contribute as
\(O(m_{\mathrm{res}}^2)\).

\section{Our Lattices}

Our action must be chosen to minimize chiral symmetry breaking while
remaining computationally fast enough to produce thousands of lattices.
Since this means our lattices will be rather coarse and our fifth
dimension relatively small, we turn to improved actions to help lower
the residual mass. Adding a rectangle term to the action accomplishes
this by smoothing the gauge field at short distances.  We use the DBW2
rectangle coefficient (\(c_1=-1.4069\)) and gauge coupling \(\beta = 0.72\)
in a gauge action of the form
\(S_g = - \frac{\beta}{3} \sum_x \left( \left(1 - 8 c_1\right) {\rm Tr}\, U_{\rm plaq} + c_1 {\rm Tr}\, U_{\rm rect}\right)\).

In summary, our lattices are \(16^3 \times 32 \times 8\) and have bare
sea quark masses \(m_s = 0.04\) and \(m_l \in (0.01,0.02)\). We used the
RHMC to generate 6000 trajectories and discarded 1000 for thermalization.
We select configurations separated by 50 for analysis, giving us a total
of 100 sets of weak matrix elements for each light sea quark mass.
Each set combines nondegenerate pairs from valence quark masses
\(m_V \in (0.01,0.02,...,0.06)\).

\section{Preliminary Results}

The rho mass determines our lattice scale in the physical sea and valence
quark mass limit ; it is consistent with the scale determined from the
static quark potential. We find \(a^{-1} = 1.6(1) \mathrm{GeV}\).

The residual mass is determined from the midpoint correlator, and agrees
with the (negative) quark mass at which the pseudoscalar mass extrapolates
to zero. We find \(m_\mathrm{res} = 0.0106(1)\), which is fairly large,
being larger than our lowest light quark mass (0.01).

The pseudoscalar mass then allows us to determine the light and
strange quark masses. We find \(m_l = 0.00171(21)\) and \(m_s = 0.042(5)\).
It is important that our strange quark mass be near our sea input
\((0.04+m_\mathrm{res})\), since we do not intend to extrapolate its value.

The pseudoscalar decay constant enters in the normalization of \BK. It
may be determined from a ratio of the wall-point pseudoscalar
correlator to the wall-wall pseudoscalar correlator.
\begin{equation}
f_P = \frac{2(m_q+m_\mathrm{res})}{M_P^2} \frac{{\cal C}^{PP}_{wp}(0,t)\sqrt{2M_P}}{{\cal C}^{PP}_{ww}(0,t)e^{-M_P t}V},
\end{equation}
where \(m_q\) and \(m_\mathrm{res}\) are quark and residual mass,
\(M_P\) is pseudoscalar meson mass, \(V\) is volume, and the \({\cal C}\)'s
are correlators with superscripts denoting the source and sink
operators (both \(P\)seudoscalar in this case) and subscripts denoting
source and sink shapes (\(p\)oint or \(w\)all).

\begin{figure}
\epsfig{file=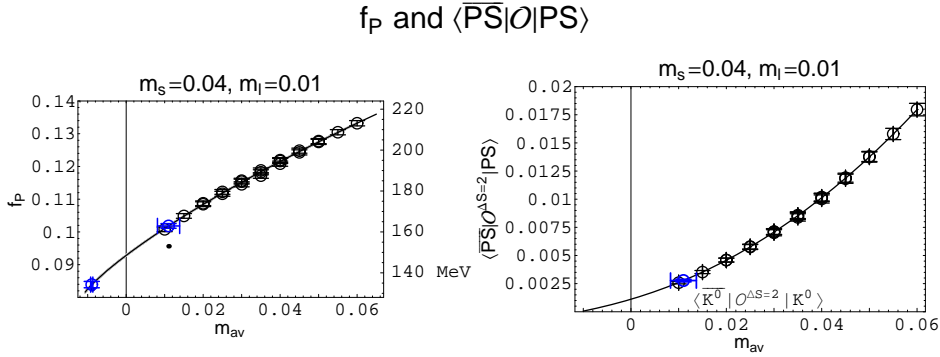,width=\textwidth}
\caption{{\bf Left}: Pseudoscalar decay constant as a function of
average valence quark mass (blue points mark physical \(f_\pi\) and \(f_K\));
{\bf Right}:
\(\left<\overline{\mathrm{PS}}|{\cal O}^{\Delta S=2}|\mathrm{PS}\right>\)
(blue point marks physical kaon mass)}
\label{fpsME}
\end{figure}

We fit \(f_P\) to the chiral form
\begin{equation}
f_P = a + b m + c m \log{m},
\end{equation}
where \(m = m_q + m_{\mathrm res}\). See Fig.~\ref{fpsME} (left).

The most naive way to derive the matrix element associated with \BK is
simply to take the three-point correlator and divide by the wall-wall
two-point correlator to remove the matrix elements associated with the
wall sources:
\begin{equation}
\left<\overline{\mathrm{PS}}|{\cal O}|\mathrm{PS}\right> = 4 M_P
   \frac{{\cal C}^{P{\cal O}P}_{wpw}(t_\mathrm{src},t,t_\mathrm{snk})}{{\cal C}^{PP}_{ww}(t_\mathrm{src},t_\mathrm{snk})}
\end{equation}
See Fig.~\ref{fpsME} (right).

However, \BK may be derived at once without using the noisy wall-wall
correlator by a clever use of wall-point correlators:
\begin{equation}
B_P = \frac{M_P^2 V}{2\frac{8}{3}(m_q+m_\mathrm{res})^2} \frac{{\cal C}^{P{\cal O}P}_{wpw}(t_\mathrm{src},t,t_\mathrm{snk})}{{\cal C}^{PP}_{wp}(t_\mathrm{src},t){\cal C}^{PP}_{wp}(t,t_\mathrm{snk})} \frac{Z_{\cal O}}{Z_A^2} Z_A^2 Z_{\overline{\mathrm{MS}}},
\end{equation}
where \(Z_A = 0.732\) is the renormalization factor of the axial
current computed on our lattices, \(Z_{\cal O}/Z_A^2 = 0.93\) is the
renormalization factor of the three-point operator
\({\cal O}^{\Delta S=2}_{LL}\), and \(Z_{\overline{\mathrm{MS}}} = 1.02\)
is the conversion factor from RI-MOM to MS-bar scheme. We take
\(Z_{\cal O}/Z_A^2\) from previous calculations on 2 flavor lattices
and \(Z_{\overline{\mathrm{MS}}}\) from previous quenched calculations.

\begin{figure}
\epsfig{file=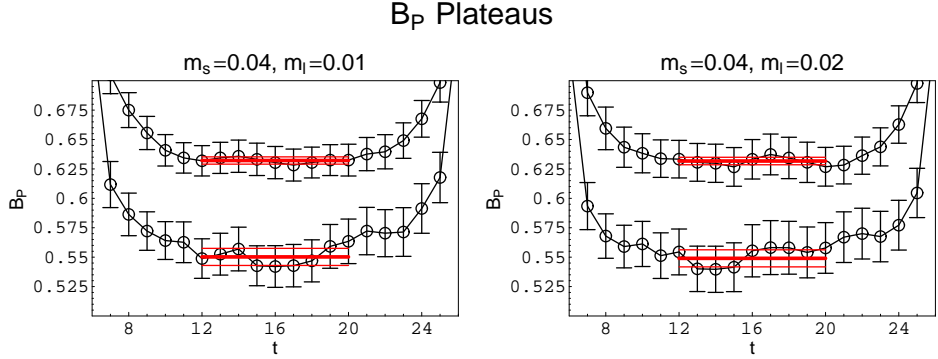,width=\textwidth}
\caption{Plateaus of \(B_P\) for highest (0.06) and lowest (0.01)
  average valence quark mass}
\label{BPplat}
\end{figure}

We expect \BK to approach its asymptotic value far from the source and
sink. Depicted in Fig.~\ref{BPplat} are the heaviest and lightest
valence quark masses (\(m_V = 0.01\) and \(0.06\)). The plateaus do
not appear to have any unusual wiggles or trends.

\begin{figure}
\epsfig{file=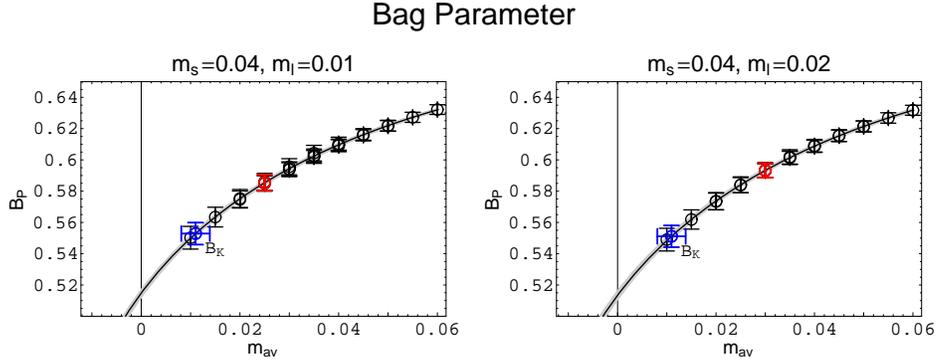,width=\textwidth}
\caption{Pseudoscalar bag parameter as a function of average valence
quark mass (blue point marks valence extrapolation to physical
\(B_K\), red point marks the valence=sea point)}
\label{BP}
\end{figure}

\BK is fit to a form including the chiral fits to \(f_P\) and \(M_P\):
\begin{equation}
\left<\overline{\mathrm{PS}}|{\cal O}|\mathrm{PS}\right> = a + b \frac{M_P^2}{4\pi f_P} \log{M_P^2} + c M_P^2,
\end{equation}
where \(M_P\) is a function of quark mass, but \(f_P\) is taken in the
chiral limit. See Fig.~\ref{BP}.

\begin{figure}
\centering
\epsfig{file=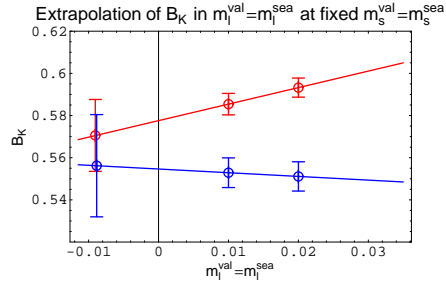,width=0.4\textwidth}
\caption{Two extrapolations of \BK to the physical point:
{\bf red:} extrapolates valence=sea to the physical point;
{\bf blue:} takes valence=physical, then extrapolates sea to the
physical point}
\label{BKextrap}
\end{figure}

Taking the valence=sea point (red on Fig.~\ref{BP}) for each set
of lattices, we may extrapolate to the chiral limit for the light
quarks (leaving the strange quarks at the physical strange quark
mass).  This yields a final value (See Fig.~\ref{BKextrap}.):
\begin{equation}
B_K = 0.571(17)
\end{equation}

Alternatively, we may take the valence=physical limit first (blue on
Fig.~\ref{BP}), taking advantage of the chiral fitting form, and then
extrapolate linearly to the sea=physical limit. This
alternate method yields a somewhat lower value of
\begin{equation}
B_K = 0.556(24)
\end{equation}

\section{Conclusions}

\begin{figure}
\centering
\epsfig{file=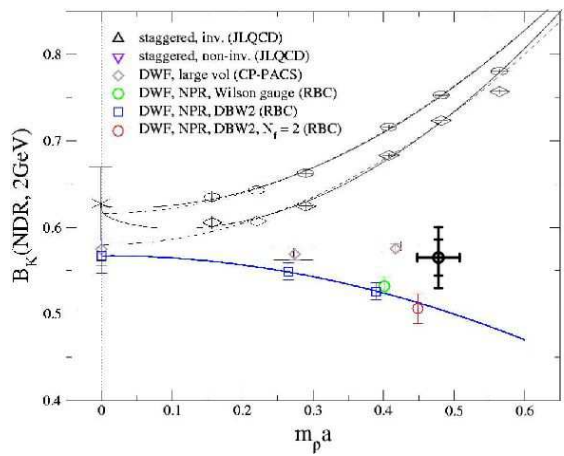,width=0.5\textwidth}
\caption{Dynamical extrapolation of \BK to the physical point of 2+1
flavor lattices at 1.6 GeV (marked in bold black) compared to other
recent values}
\label{BKmerge}
\end{figure}

Since neither extrapolation correctly accounts for nonlinear chiral
terms, we take the average of the two and add a systematic error
associated with the difference:
\begin{equation}
B_K = 0.563(15)(21),
\end{equation}
where the first error is systematic and the second statistical.  In
either case, the preliminary 2+1 value of \BK is somewhat higher than
the scale extrapolation from 2 flavor DWF values. See Fig.~\ref{BKmerge}.

In the future, we wish to improve this calculation with
larger volumes, smaller residual masses and a full nonperturbative
treatment of renormalization. Larger volumes will allow us to use
longer plateaus to reduce statistical error and also diminish any
finite-volume effects. Smaller residual masses will further diminish
mixing with wrong-chirality operators. A correct treatment of the
factors \(Z_{\cal O}\) and \(Z_{\overline{\mathrm{MS}}}\) on 2+1
flavor lattices is necessary to complete this calculation, which will
require NPR techniques.

\section*{Acknowledgements}

We thank Sam Li, Meifeng Lin, Chris Maynard and Robert Tweedie for
help generating datasets. We thank Peter Boyle, Dong Chen, Norman
Christ, Mike Clark, Calin Cristian, Zhihua Dong, Alan Gara, Andrew
Jackson, Balint Joo, Chulwoo Jung, Richard Kenway, Changhoan Kim,
Ludmila Levkova, Xiaodong Liao, Guofeng Liu, Shigemi Ohta, Konstantin
Petrov, Tilo Wettig and Azusa Yamaguchi for developing with us the
QCDOC hardware and software. This development and the resulting
computer equipment used in this calculation were funded by the
U.S. DOE grant DE-FG02-92ER40699, PPARC JIF grant PPA/J/S/1998/00756
and by RIKEN.  This work was supported by DOE grant DE-FG02-92ER40699
and we thank RIKEN, BNL and the U.S. DOE for providing the facilities
essential for the completion of this work.

\end{document}